\documentclass[10pt,twocolumn,twoside]{IEEEtran}

\usepackage{graphicx}
\usepackage{amsmath}
\usepackage{empheq}
\usepackage{amsfonts}
\usepackage{amssymb}
\usepackage{amsthm}
\usepackage[labelformat=simple]{subcaption}
\usepackage{citesort}
\usepackage{varwidth}
\usepackage{algpseudocode}
\usepackage{algorithm}
\newcommand{\subparagraph}{}
\usepackage{titlesec}
\usepackage{placeins}
\usepackage[normalem]{ulem}
\usepackage{color}
\usepackage{multirow}
\usepackage{xcolor}
\usepackage{url}

\titlespacing\section{0pt}{8pt plus 1pt minus 1pt}{4pt plus 2pt minus 2pt}
\titlespacing\subsection{0pt}{7pt plus 1pt minus 1pt}{4pt plus 1pt minus 1pt}
\titlespacing\subsubsection{0pt}{4pt plus 1pt minus 1pt}{4pt plus 1pt minus 1pt}
\titlespacing\paragraph{0pt}{7pt plus 1pt minus 1pt}{4pt plus 1pt minus 1pt}

\usepackage{listings}

\definecolor{myorange}{rgb}{1,0.4,0}

\lstdefinestyle{customnesc}{
  belowcaptionskip=1\baselineskip,
  breaklines=true,
  frame=single,
  xleftmargin=\parindent,
  captionpos=b,
  language=C,
  showstringspaces=false,
  basicstyle=\footnotesize\ttfamily,
  commentstyle=\itshape\color{green!40!black},
  identifierstyle=\color{black},
  stringstyle=\color{myorange},
  classoffset=0,
  keywordstyle=\bfseries\color{blue},
  morekeywords={event, call, uint8_t},
  classoffset=1,
  morekeywords={radio_msg_t, serial_msg_t, message_t},
  keywordstyle=\color{black},
  classoffset=0,
}

\lstdefinestyle{customcsharp}{
  belowcaptionskip=1\baselineskip,
  breaklines=true,
  frame=single,
  xleftmargin=\parindent,
  captionpos=b,
  language=C++,
  showstringspaces=false,
  commentstyle=\itshape\color{green!40!black},
  stringstyle=\color{myorange},
  basicstyle=\footnotesize\ttfamily,
  keywordstyle=\color{blue}\bfseries,
  morekeywords={ushort, ulong, byte},
}

\begin{document}

\color{black}

\title{\centering Hybrid-BCP: A Robust Load Balancing and Routing Protocol for Intra-Car Wired/Wireless Networks}

\renewcommand{\footnoterule}{%
  \kern -3pt
  \hrule width \textwidth height 0pt
  \kern 3pt
}


\author{\IEEEauthorblockN{Wei Si\textsuperscript*, David Starobinski\textsuperscript*, and Moshe Laifenfeld\textsuperscript{\dag} }
\IEEEauthorblockA{\\\textsuperscript*Dept. of Electrical and Computer Engineering,
Boston University, USA\\ }
\IEEEauthorblockN{\{weisi, staro\}@bu.edu\\}
\IEEEauthorblockN{\textsuperscript{\dag}moshel@spacegate.com}\vspace*{-0.8cm}
}

\maketitle

\begin{abstract}

With the emergence of connected and autonomous vehicles, sensors are increasingly deployed within cars to support new functionalities. Traffic generated by these sensors congest traditional intra-car networks, such as CAN buses. Furthermore, the large amount of wires needed to connect sensors makes it harder to design cars in a modular way. To alleviate these limitations, we propose, simulate, and implement a \emph{hybrid} wired/wireless architecture, in which each node is connected to either a wired interface or a wireless interface or both. Specifically, we propose a new protocol, called \emph{Hybrid-Backpressure Collection Protocol (Hybrid-BCP)}, to efficiently collect data from sensors in intra-car networks. Hybrid-BCP is backward-compatible with the CAN bus technology, and builds on the BCP protocol, designed for wireless sensor networks. Hybrid-BCP achieves high throughput and shows resilience to dynamic network conditions, including adversarial interferences. Our testbed implementation, based on CAN and ZigBee transceivers, demonstrates the load balancing and routing functionalities of Hybrid-BCP and its resilience to DoS attacks. We further provide simulation results, obtained with the ns-3 simulator and based on real intra-car RSSI traces, that compare between the performance of Hybrid-BCP and a tree-based data collection protocol. Notably, the simulations show that Hybrid-BCP can achieve the same performance as the tree-based protocol while reducing the radio transmission power by a factor of 10.


\end{abstract}

\section{Introduction}

The conventional intra-car communication model, in which sensors  communicate with Electronic Control Units (ECUs) via  CAN buses, faces several limitations.
First, the increasing amount of wires required to connect sensors to the intra-car network results in fuel inefficiency and complicates car design and maintenance \cite{Morteza2013}.
Second, new sensors generates additional traffic and increases the likelihood of congestion on the CAN bus.
Last, since the CAN protocol is based on message priority, it is vulnerable to Denial-of-Service (DoS) attacks generated by high-priority messages \cite{Koscher-2010}.

To alleviate limitations of intra-car wired networks, we propose in this work a hybrid wired/wireless network architecture for supporting intra-car communication.
A key goal in this context is to achieve reliable and efficient delivery of packets from the sensors to a sink (ECU), a task also known as \emph{data collection}.

\begin{figure}[t!]
\centering
\includegraphics[scale=0.35, clip=true, trim=1.28cm 3.70cm 1.15cm 3.35cm]{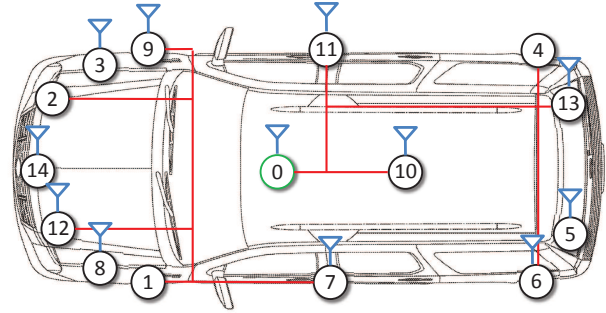}
\caption{A 15-node intra-car hybrid wired/wireless network. Each node is connected to either a wired interface or a wireless interface or both. The data packets of the sensor nodes (1-14) need to be delivered to the sink (node 0).}
\label{fig:hybrid_network_illustrated}
\vspace*{-0.3cm}
\end{figure}

The design of such a hybrid network brings up several research issues. The first issue is how to implement routing. For instance, in the hybrid network of Fig. \ref{fig:hybrid_network_illustrated},  packets destined from node~2 to the sink can be routed either through node 7 or node 9. Which node should be chosen as the next hop?

The second issue is how to implement load balancing. For instance, node 10 can communicate with the sink either on the wired interface or the wireless interface. Which interface should be used?

The third issue is how to deal with contention from other nodes and (possibly malicious)  interferences.  For instance, how should node 10 react if node 4 is contending on the wired link? And what happens if an adversary performs a DoS attack?


In light of these challenges, we define the following objectives for designing a collection protocol for hybrid intra-car networks:
\begin{itemize}

\item \emph{Load balancing.} The protocol should balance packet transmissions over available interfaces.

\item \emph{Routing.} In the absence of a direct communication link between a sensor node and the sink, the protocol should deliver the packets of the sensor node in a multi-hop fashion.


\item \emph{Robustness.} The protocol should achieve reliable data collection even when link qualities degrade (e.g., due to contention, interferences, or DoS attacks).
    
\item \emph{Backward-compatibility} The protocol should not require the replacement of existing technology (e.g., CAN buses) in vehicles.

\end{itemize}


Routing protocols based on the construction and maintenance of end-to-end routes adapt poorly to intra-car wireless channel conditions that typically experience deep fading and high variability \cite{bcp}.

To address these issues, we propose a new data collection protocol for hybrid intra-car networks, called \emph{Hybrid-BCP}. Hybrid-BCP belongs to the class of \emph{backpressure} algorithms \cite{bcp}, which have theoretically been proven to be throughput-optimal. 

Hybrid-BCP does not calculate end-to-end routes. Rather it relies on a distributed computation of \emph{backpressure weights}. Each node maintains a backpressure weight on each interface for each of its neighbors, based on the link quality and the differential of the queue lengths. For each incoming packet, a node selects the interface/neighbor combination with the highest positive backpressure weight as the next hop. If all the backpressure weights are negative, then the node stores the packet in its queue and waits until one of the backpressure weights becomes positive.
 
We implement Hybrid-BCP on a real testbed, composed of CAN and ZigBee transceivers, and evaluate its performance. Our testbed experiments demonstrate the load balancing and routing functionalities of Hybrid-BCP. The results show that Hybrid-BCP improves throughput under DoS attacks on the CAN bus by a factor of 10. 
They also show that Hybrid-BCP is robust to jamming attacks on wireless links.

We further implement Hybrid-BCP in ns-3 for the purpose of simulating a larger network. We compare Hybrid-BCP with a tree-based collection protocol, which we refer to as \emph{Hybrid-Collection Tree Protocol (Hybrid-CTP)}. Hybrid-CTP relies on the computation and update of end-to-end routing metrics at each node.

For the simulations, we use real RSSI (received signal strength indication) traces collected in an intra-car environment \cite{teacp-2015}.
The simulation results demonstrate that Hybrid-BCP achieves higher reliability than Hybrid-CTP if both protocols use the same power transmission (e.g., 95\% vs 88\%). Conversely, Hybrid-BCP can reduce the radio transmission power by a factor of 10 and still achieve the same reliability as Hybrid-CTP.

We summarize the contributions of this paper as follows:

\begin{itemize}

\item We design a new protocol, Hybrid-BCP, for data collection in intra-car hybrid wired/wireless networks.

\item We build a real testbed for evaluating the performance of Hybrid-BCP. The tests demonstrate the load balancing and routing functionalities of Hybrid-BCP and its resilience to DoS attacks.

\item We implement Hybrid-BCP and Hybrid-CTP in the ns-3 simulator, and compare their performance in terms of reliability for different transmission powers.

\end{itemize}

The rest of the paper is organized as follows. Section \ref{sec:related_work} reviews related work on hybrid wired/wireless networks, load balancing algorithms for multiple interfaces, and collection protocols. Section \ref{sec:hybrid-bcp} describes the Hybrid-BCP protocol and its software implementation. Section \ref{sec:experiments} and \ref{sec:simulations} provide performance evaluation of Hybrid-BCP in testbed experiments and simulations, respectively. Finally, Section \ref{sec:conclusion} concludes the paper and discusses future research directions.

\section{Related work} \label{sec:related_work}

\subsection{Hybrid wired/wireless networks}

Much of the existing work on hybrid wired/wireless networks assumes that all the devices (except for bridges or relays) are connected to either a wired interface or a wireless interface but not both.

For instance, \cite{Mirabella-2011} implements a hybrid wired/wireless network for greenhouse control and management using CAN and ZigBee transceivers. In that system, the central controller and a number of wireless bridges are connected to a bus. The bridges receive data from wireless sensors and forward them to the controller. The work in \cite{Miorandi-2004} conducts a feasibility study of a hybrid wired/wireless network implementation based on Ethernet and Bluetooth. In the implementation, sensors have either a wired or a wireless interface while the sinks are connected to a bus. A bridge node communicates between the wireless nodes and the wired nodes. Similar hybrid network structures can be found in \cite{sun-2003,seno-2011}, where wireless nodes communicate with wired nodes through access points. 
In the hybrid wired/wireless models of \cite{Liu-2003,Zemlianov-2005}, a number of base stations are interconnected with high-bandwidth wired links and they serve as relays for the wireless nodes. 

In contrast to the above work, Hybrid-BCP allows any node (sensors and ECUs) to be connected to either type of interfaces or both.


\subsection{Load balancing}

There exist several protocols for aggregating bandwidth and performing end-to-end load balancing. These protocols are implemented at the transport layer or above, and rely on protocols at lower layers to provide the routing functionality.

For instance, Multipath TCP (MPTCP)~\cite{MPTCP} uses multiple TCP paths to increase the throughput of data transfer.  The earliest delivery path first (EDPF)~\cite{EDPF} estimates the packet delivery time on severals path and schedules packets on the path with the shortest delivery time. The work in~\cite{liu-2007} adds to EDPF by incorporating transmission rates and losses in the estimation of the delivery time of packets. Other algorithms based upon EDPF includes~\cite{Fernandez-2009,Chebrolu-2005,Ramaboli-2010}.

Different from the above work, Hybrid-BCP provides a joint load balancing and routing solution.

\subsection{Collection protocols}

\begin{algorithm}[t]
\caption{BCP}
\begin{varwidth}{\dimexpr\linewidth-2\fboxsep-2\fboxrule\relax}
\begin{algorithmic}[1]
\State Compute backpressure weight $w_{i,j}$ for each neighbor $j$
\State Find the neighbor $j^*$ such that $j^*=\operatorname*{arg\,max}_j w_{i,j}$
\If {$w_{i,j*}$ $>0$}
\State Transmit a packet to $j^*$
\State Update $\overline{ETX}_{i\rightarrow j^*}$ and $\overline{\mathcal{R}}_{i\rightarrow j^*}$
\Else
\State Wait for a reroute period and go to line 1
\EndIf
\State Go to line 1
\end{algorithmic}
\end{varwidth}%
\end{algorithm} 

Collection protocols are routing protocols designed specifically for routing data from sensor nodes to a central collection node. There exist two well-known collection protocols in wireless sensor networks. The first one is the Collection Tree Protocol (CTP)~\cite{ctp-journal}. CTP establishes a minimum-cost routing tree where the cost on each link cost equals the expected number of transmissions on that link (ETX).

The other one is the Backpressure Collection Protocol (BCP)~\cite{bcp}. BCP derives from backpressure routing algorithms, which achieve optimal throughput. With BCP, nodes independently make routing decisions based on local information. Routing decisions are made on a per packet basis rather than on a per-computed path. The work in \cite{bcp} shows that BCP achieves higher throughput and reliability than CTP  under dynamic network conditions (e.g., in the presence of external sources of interferences).

Since Hybrid-BCP is built upon BCP, we next briefly review how BCP makes routing decisions. Let $\mathcal{Q}_i$ represent the backlog (i.e., number of packets stored) at node $i$. The $\Delta\mathcal{Q}_{i,j}=\mathcal{Q}_i -\mathcal{Q}_j$ is the queue differential (backpressure) between node $i$ and its neighbor node $j$. Let  $\overline{\mathcal{R}}_{i\rightarrow j}$ be the estimated link rate from $i$ to $j$ and let $\overline{ETX}_{i\rightarrow j}$ be an estimate of the average number of transmissions needed to successfully transmit a packet over the link. According to the routing policy of BCP, node $i$ calculates the backpressure weight for each neighbor $j$ as follows:
\[
w_{i,j} = (\Delta \mathcal{Q}_{i,j} - V\cdot \overline{ETX}_{i\rightarrow j}) \cdot \overline{\mathcal{R}}_{i\rightarrow j}.
\]

The routing decision (i.e., the selected next hop for the current packet) is determined by finding the neighbor $j^*$ with the highest weight. Node $i$ then makes the forwarding decision: if $w_{i,j^*}>0$, then the packet is forwarded to node $j^*$, \emph{else the packet is held until the metric is recomputed}. In other words, if the weights for all neighbor nodes are zero or negative, the node will do nothing but wait until the next recomputation (after a \textit{reroute period}). A pseudo-code of BCP is given in Algorithm 1.

BCP estimates $\overline{ETX}$ based on an exponential moving weighted average formula. Whenever a new sample of $ETX$ is obtained, $\overline{ETX}$ is updated as follows: $\overline{ETX}_{new} = \alpha\overline{ETX}_{old} + (1-\alpha)ETX$. The default value of $\alpha$ is 0.9. The link rate is calculated as the reciprocal of the packet transmission time (the time elapsing from the first transmission to the reception of an ACK), and the estimated link rate $\overline{\mathcal{R}}$ is updated according to an exponential moving weighted average formula similar to that used for $\overline{ETX}$.

\section{Hybrid-BCP} \label{sec:hybrid-bcp}

In this section, we describe the protocol design of Hybrid-BCP and its software implementation.

\subsection{Protocol design}

Hybrid-BCP can be viewed as two BCP algorithms running in parallel, with one algorithm handling the wired interface (e.g., CAN) and the other one handling the wireless interface (e.g., ZigBee).

Next, we describe the handler of interface $I$, where $I \in \{W, WL\}$ ($W$ represents the wired interface and $WL$ represents the wireless interface). 
Let  $\overline{\mathcal{R}}^{I}_{i\rightarrow j}$ be the estimated link rate from $i$ to $j$ over interface $I$ and let $\overline{ETX}^{I}_{i\rightarrow j}$ be an estimate of the average number of transmissions needed to successfully transmit a packet over the interface. The interface handler of node $i$ calculates the following backpressure weight for each neighbor $j$ on interface $I$ as follows:
\[
w_{i,j}^{I} = (\Delta \mathcal{Q}_{i,j} - V\cdot \overline{ETX}^{I}_{i\rightarrow j}) \cdot \overline{\mathcal{R}}^{I}_{i\rightarrow j}.
\]
Let $j^*$ denote the neighbor with the highest weight on the wired interface, i.e., $j^* = \operatorname*{arg\,max}_{j} w_{i,j}^{W}$. Let $k^*$ denote the neighbor with the highest weight on the wireless interface, i.e., $k^* = \operatorname*{arg\,max}_{k} w_{i,k}^{WL}$.

\begin{algorithm}[t!]
\caption{Hybrid-BCP}
\begin{varwidth}{\dimexpr\linewidth-2\fboxsep-2\fboxrule\relax}
\begin{algorithmic}[1]
\Procedure{Wired\_interface\_handler}{}
\While{$\mathcal{Q}_i > 0$}
\State Wire\_busy $\leftarrow$ false
\State Compute the backpressure weight $w_{i,j}^{W}$ for \hspace*{1.05cm}each neighbor $j$ on the wired link
\State Find the neighbor $j^*$ such that $j^*=$ \hspace*{1.05cm}$\operatorname*{arg\,max}_{j} w_{i,j}^{W}$
\If {$w_{i,j^*}^{W}>0$ \textbf{and} (Wireless\_busy = true \textbf{or} \hspace*{1.3cm} $w_{i,j^*}^{W} \geq w_{i,k^*}^{WL}$)}
\State Wire\_busy $\leftarrow$ true
\State Transmit one packet to $j^*$ over the wired \hspace*{1.57cm}interface
\State Update $\overline{ETX}^{W}_{i\rightarrow j^*}$ and $\overline{\mathcal{R}}^{W}_{i\rightarrow j^*}$
\State Wire\_busy $\leftarrow$ false
\Else
\State Wait for a reroute period 
\EndIf
\EndWhile
\EndProcedure
\State 	
\Procedure{Wireless\_interface\_handler}{}
\While{$\mathcal{Q}_i > 0$}
\State Wireless\_busy $\leftarrow$ false
\State Compute the backpressure weight $w_{i,k}^{WL}$ for \hspace*{1.05cm}each neighbor $k$ on the ZigBee links
\State Find the neighbor $k^*$ such that $k^*=$ \hspace*{1.05cm}$\operatorname*{arg\,max}_{k} w_{i,k}^{WL}$
\If {$w_{i,k^*}^{WL}>0$ \textbf{and} (Wire\_busy = true \textbf{or} \hspace*{1.3cm} $w_{i,k^*}^{WL} \geq w_{i,j^*}^{W}$)}
\State Wireless\_busy $\leftarrow$ true
\State Transmit one packet to $k^*$ over the wireless \hspace*{1.57cm}interface
\State Update $\overline{ETX}^{WL}_{i\rightarrow k^*}$ and $\overline{\mathcal{R}}^{WL}_{i\rightarrow k^*}$
\State Wireless\_busy $\leftarrow$ false
\Else
\State Wait for a reroute period 
\EndIf
\EndWhile
\EndProcedure
\end{algorithmic}
\end{varwidth}%
\end{algorithm}


A higher backpressure weight represents a link of higher quality and a neighbor with less backlog. A necessary condition for the wired interface handler to transmit a packet to neighbor $j^*$ is that $w_{i,j^*}^{W} > 0$. When both the wired and wireless interface handlers are idle, an additional condition is that the weight of the wired interface is the larger one, i.e., $w_{i,j^*}^{W} \geq w_{i,k^*}^{WL}$. If one of these conditions is not satisfied, then the wired interface handler waits for the next computation of backpressure weights. Similar conditions apply for the wireless interface handler. Algorithm 2 provides a pseudo-code of Hybrid-BCP.  Table~\ref{tab:scheduling_hybrid-bcp} summarizes the scheduling procedure of Hybrid-BCP.

\begin{table}[t!]
\centering
\begin{tabular}{|c|c|p{5cm}|}
\hline
$w_{i,j^*}^{W}$ & $w_{i,k^*}^{WL}$ & Operation \\ \hline
$> 0$ & $\leq 0$ & Transmit the next packet to neighbor $j^*$ on the wired link.\\ \hline
$\leq 0$ & $> 0$ & Transmit the next packet to neighbor $k^*$ on the wireless link. \\ \hline
$\leq 0$ & $\leq 0$ & The next packet is not transmitted. \\ \hline
$> 0$ & $> 0$ & If both interface handlers are idle, the next packet is scheduled on the link with the larger weight. If one of the interface handlers is busy, the next packet is transmitted on the interface which is idle. \\ \hline
\end{tabular}
\vspace*{0.3cm}
\caption{Packet transmission scheduling of Hybrid-BCP.}
\label{tab:scheduling_hybrid-bcp}
\end{table}

\begin{figure}[t]
\centering
\includegraphics[scale=0.50, clip=true, trim=0cm 0cm 8.71cm 7.59cm]{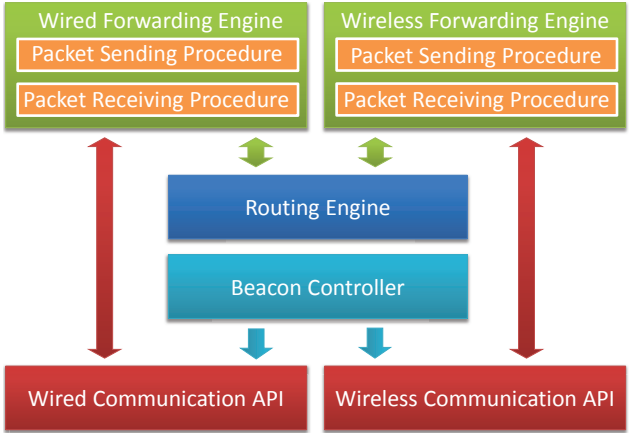}
\caption{The software architecture of Hybrid-BCP.}
\label{fig:hybrid-bcp-arch}
\vspace*{-0.3cm}
\end{figure}

\subsection{Software implementation}

The software implementation of Hybrid-BCP consists of a routing engine, a wired forwarding engine, a wireless forwarding engine and a beacon controller (see Fig. \ref{fig:hybrid-bcp-arch}). 

The routing engine is responsible for calculating the backpressure weights for each neighbor and interface. It updates and maintains the routing table. 

The forwarding engine is responsible for scheduling packet transmissions and handling packet receptions. It is further composed of a packet sending procedure and a packet receiving procedure: the packet sending procedure runs the interface handler described in Algorithm 2, while the packet receiving procedure handles ACK packets and provides information for the routing engine to update the routing table. 

The forwarding engine also keeps a count of transmissions for each packet. When the packet sending procedure transmits a packet on the interface, it waits to receive an ACK from the next hop until an \emph{ACK timeout}. If an ACK is not received before the timeout, the packet sending procedure retransmits the packet on the interface. 

Hybrid-BCP utilizes \emph{beacon messages} to propagate backpressure information from a node to its neighbors. The beacon controller is responsible for broadcasting beacon messages on all available interfaces.


\section{Experiments}
\label{sec:experiments}

In this section, we demonstrate the load balancing and routing functionalities of Hybrid-BCP in the testbed. We also show that Hybrid-BCP can be used to protect against DoS attacks on the CAN bus and wireless jamming attacks.

\subsection{Performance metrics}

Before presenting the experiments, we provide the definition of metrics for evaluating the performance of Hybrid-BCP. 

Suppose a test lasts for $T$ seconds. Let $N$ denote the total number of generated packets. Let $N_u$ denote the number of delivered packets, excluding packet duplicates, and let $\mathcal{S}_d$ represent the set of the uniquely delivered packets.

The \emph{delivery rate} is defined to be the percentage of packets that are delivered, i.e., ${N_u \over N} \cdot 100\%.$ The \emph{throughput} is defined to be the number of unique packets delivered to the sink per second, i.e., $\frac{N_u}{T} \text{ pkts/sec}.$ The delay of a packet $D_i$ is defined as the time elapsing from its generation at the source node to its delivery at the sink. The \emph{average delay} is calculated as ${1\over N_u} \sum_{i \in \mathcal{S}_d} D_i.$

\subsection{Experimental setup}

\begin{table}[t!]
\centering
\begin{tabular}{|c|c|}
\hline
ACK timeout for CAN link & 30 ms \\ \hline
ACK timeout for ZigBee link & 80 ms \\ \hline
Reroute period & 50 ms \\ \hline
Beaconing period & 1500-2000 ms \\ \hline
Queue size & 48 \\ \hline
\end{tabular}
\vspace*{0.3cm}
\caption{Parameters in the implementation of Hybrid-BCP for the testbed.}
\vspace*{-0.5cm}
\label{tab:testbed_parameters}
\end{table}

We build a hybrid CAN/ZigBee network to test Hybrid-BCP. We use VN1610 CAN interfaces \cite{VN1610}, manufactured by Vector Informatik GmbH, as CAN transceivers. We use TelosB motes \cite{telosb} as ZigBee transceivers. The CAN bus is configured to operate at the rate of 33,333 baud. The transfer rate of a ZigBee transceiver is 250 Kb/s.

To emulate a node (a sensor or an ECU), we use a laptop to which one or both types of transceivers are connected. The laptop runs a Windows Presentation Foundation (WPF) application \cite{WPF} to manage the interfaces. Hybrid-BCP is implemented in C\texttt{\#}, as a component of the WPF application.

\begin{figure}[t!]
\captionsetup[subfigure]{labelformat=empty}
\centering
\begin{subfigure}[t]{0.24\textwidth}
\centering
\includegraphics[scale=0.35, clip=true, trim=6.45cm 5.67cm 7.26cm 5.10cm]{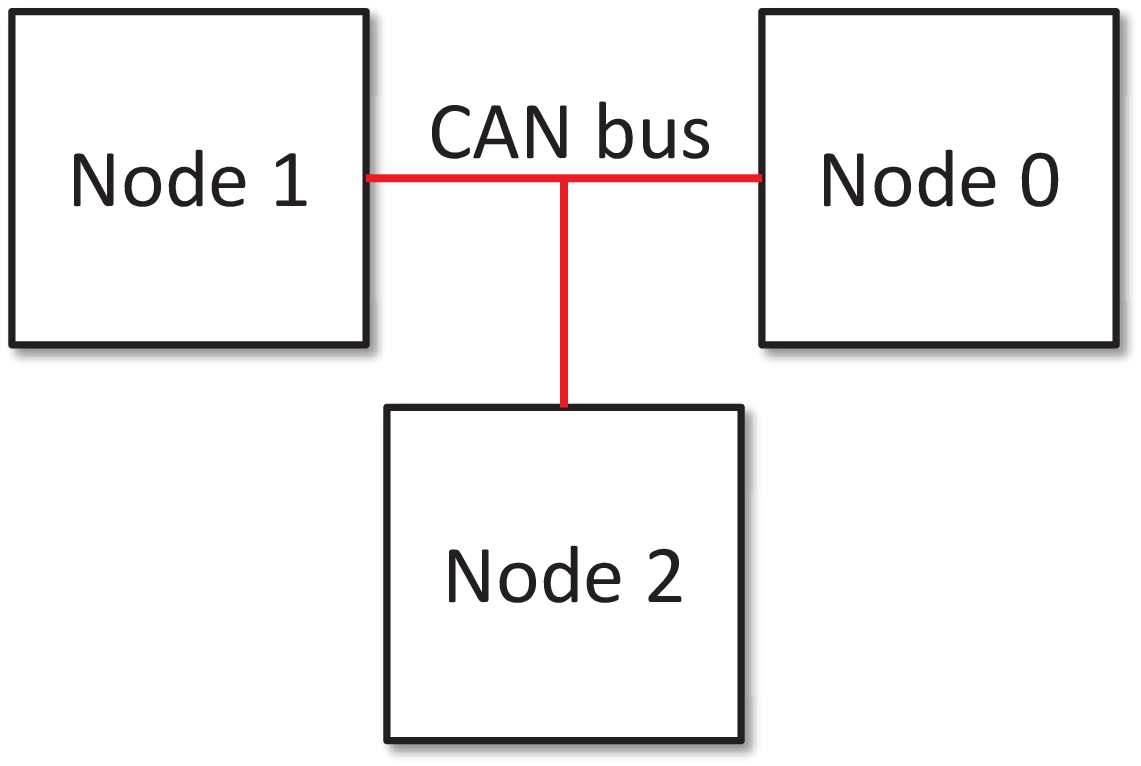}
\caption{Network A}
\end{subfigure}
\begin{subfigure}[t]{0.24\textwidth}
\centering
\includegraphics[scale=0.35, clip=true, trim=6.45cm 4.82cm 7.26cm 4.14cm]{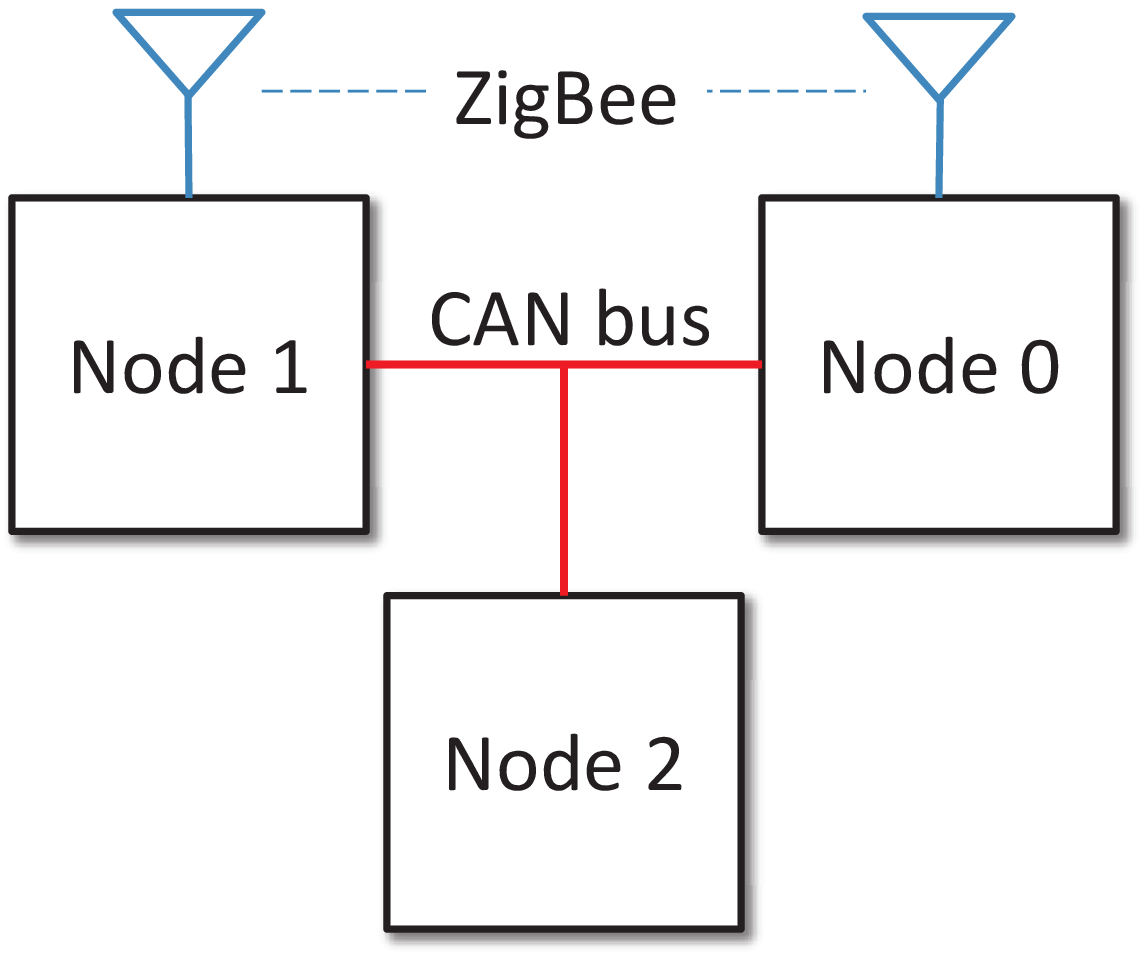}
\caption{Network B}
\end{subfigure}
\begin{subfigure}[t]{0.4\textwidth}
\centering
\includegraphics[scale=0.35, trim=3.06cm 8.86cm 4.11cm 4.13cm, clip=true]{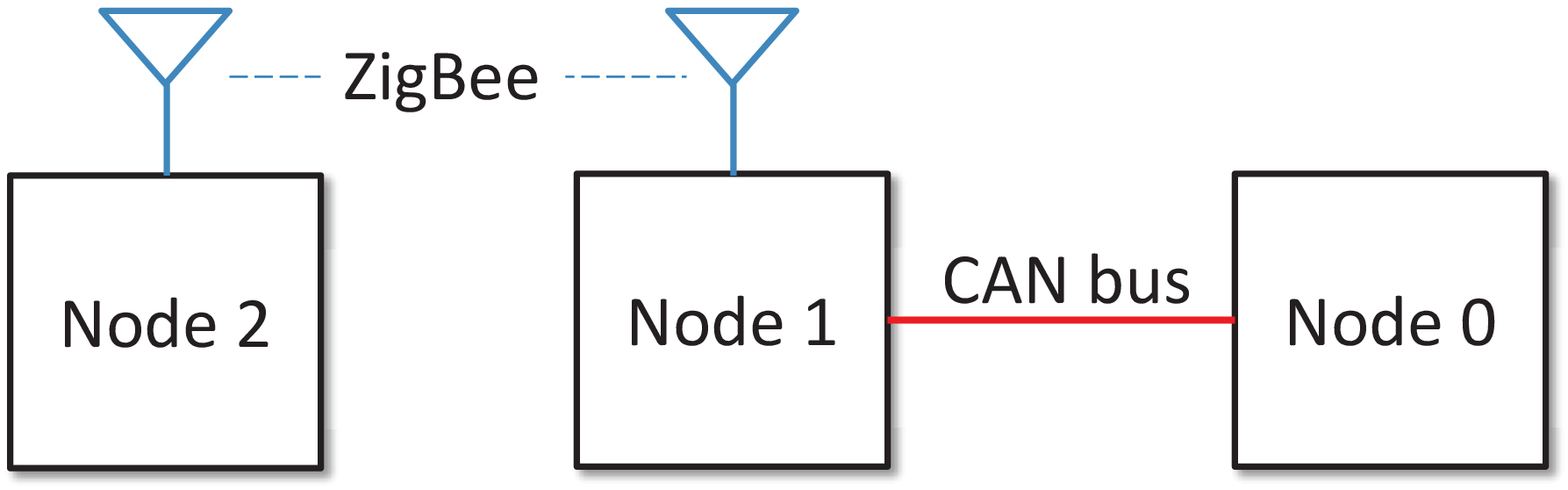}
\caption{Network C}
\label{fig:three_node_chain_network}
\end{subfigure}
\caption{The network topologies used for demonstrating the load balancing and routing functionalities of Hybrid-BCP on the testbed.}
\label{fig:topologies_load_balancing}
\vspace*{-0.3cm}
\end{figure}

\begin{figure}[t!]
\captionsetup[subfigure]{labelformat=empty}
\centering
\includegraphics[scale=0.35, trim=1.24cm 2.1cm 1.3cm 4.8cm, clip=true]{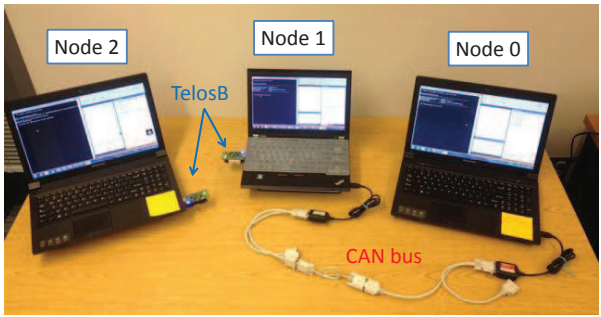}
\caption{Testbed setup for network C.}
\label{fig:testbed_setup}
\vspace*{-0.3cm}
\end{figure}

The first set of tests is conducted on the networks A, B, and C, whose topologies are shown in Fig. \ref{fig:topologies_load_balancing}. Fig. \ref{fig:testbed_setup} shows the testbed setup of network C.

We choose the ACK timeout for a CAN/ZigBee link to be slightly larger than the round trip time (RTT) of the link under light load conditions. The RTT of a CAN link is around 15 ms and that of a ZigBee link ranges from 50 ms to 70 ms. The ZigBee link has a higher RTT than a CAN link because ZigBee is based on CSMA/CA (Carrier Sense Multiple Access / Collision Avoidance) while CAN is based on CSMA/CD (Carrier Sense Multiple Access / Collision Detection).




Every time a beacon packet is transmitted, Hybrid-BCP waits for a \emph{beaconing period} to transmit a new beacon packet. The beaconing period is chosen to be sufficiently large so that beacon packets do not cause congestion on the links. It is also uniformly randomly selected within a range of possible values to avoid possible synchronization of beacon packets between different nodes and contention on the links. Table \ref{tab:testbed_parameters} lists the parameters used in the Hybrid-BCP implementation.

In the tests, each sensor node periodically generates data packets and transfers them to Hybrid-BCP, which delivers the packets to the sink. Sensor nodes generate packets at the same rate. Each test is run for five times to obtain an average and a 95\% confidence interval for the metrics. Each run lasts from three to five minutes.



\subsection{Load balancing}

\begin{figure}[t!]
\begin{subfigure}[t]{0.48\textwidth}
\centering
\includegraphics[scale=0.4]{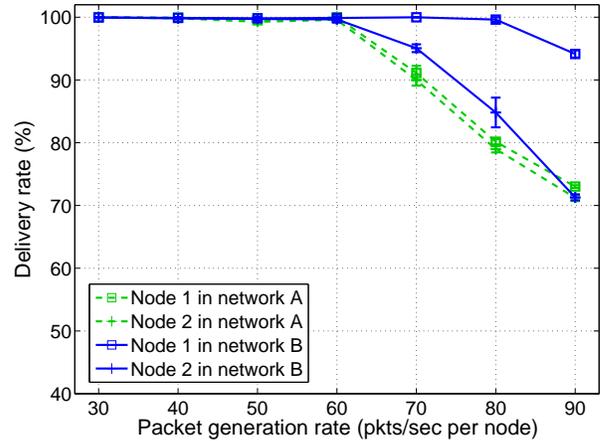}
\caption{Delivery rate versus packet generation rate.}
\label{fig:delivery_rate_three_node}
\end{subfigure}
\begin{subfigure}[t]{0.48\textwidth}
\centering
\includegraphics[scale=0.4]{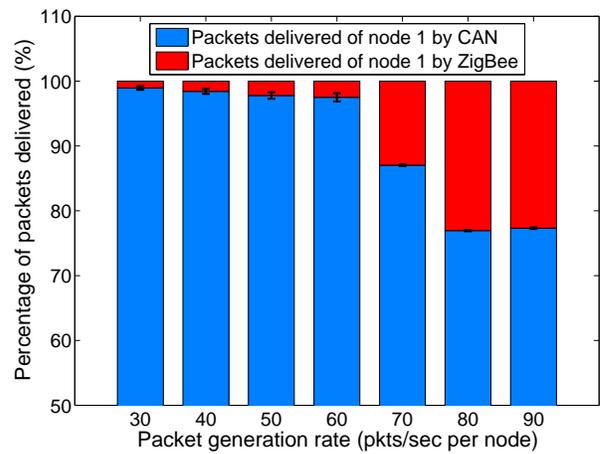}
\caption{Percentage of packets delivered over the CAN/ZigBee interfaces on network B.}
\label{fig:CAN_percentage_three_node_hybrid}
\end{subfigure}
\caption{Performance of Hybrid-BCP on network A (CAN only) and network B (hybrid).}
\label{fig:performance_three_node}
\vspace*{-0.3cm}
\end{figure}

To demonstrate the load balancing functionality of Hybrid-BCP, we perform tests on network A (a CAN network) and network B (a hybrid CAN/ZigBee network). 




Fig. \ref{fig:delivery_rate_three_node} shows that at a packet generation rate of 80 pkts/sec, Hybrid-BCP improves the delivery rate of node 1 from 80.15\% to 99.63\% thanks to the additional wireless link. Moreover, the delivery rate of node 2 also improves, from 78.99\% to 84.82\%. This is because Hybrid-BCP transmits a fraction of packets of node 1 on the ZigBee link for the purpose of load balancing, hence reducing MAC contention on the CAN bus.

In network B, when the packet generation rate of node 1 is low, Hybrid-BCP schedules most of its packets on the CAN interface, as shown by Fig. \ref{fig:CAN_percentage_three_node_hybrid}. This is because the CAN link has a smaller RTT and thus a higher link rate than the ZigBee link. When the packet rate increases, the backlog of node 1 grows and Hybrid-BCP starts scheduling packet transmissions on the ZigBee link. When the packet rate reaches a certain threshold, the percentages of packets delivered through the CAN and ZigBee interfaces do not change further because both links are saturated.

\subsection{Routing}

\begin{figure}[t!]
\includegraphics[scale=0.4]{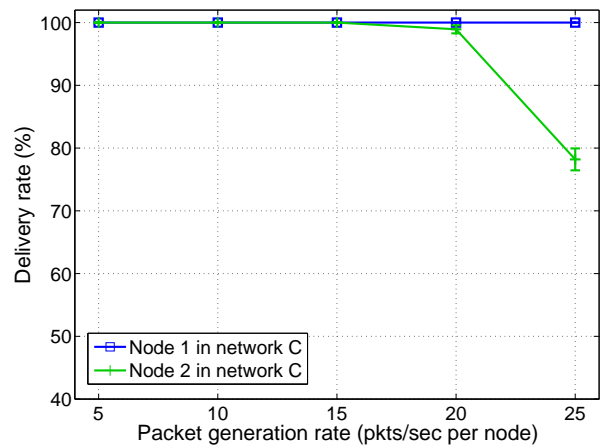}
\caption{Delivery rate versus packet generation rate of Hybrid-BCP on network C. Hybrid-BCP can successfully route the packets of node 2 to the sink via node 1. }
\label{fig:delivery_rate_three_node_chain}
\vspace*{-0.3cm}
\end{figure}

To demonstrate the routing functionality of Hybrid-BCP, we perform tests on network C. In network C, there is no direct communication link between node 2 and the sink. The packet delivery rates of node 1 and node 2 are plotted in Fig. \ref{fig:delivery_rate_three_node_chain}. The results show that the delivery rate of node 2 is 98.93\% at a packet generation rate of 20 pkts/sec. Thus, Hybrid-BCP can successfully route packets from node 2 to the sink via node 1. The ns-3 simulations in Section \ref{sec:simulations} demonstrate the routing functionality of Hybrid-BCP in a larger hybrid network.

\subsection{Robustness}

\subsubsection{DoS attacks on CAN}

\begin{figure}[t!]
\centering
\begin{subfigure}[t]{0.24\textwidth}
\centering
\includegraphics[scale=0.33, clip=true, trim=6.24cm 8.03cm 7.36cm 2.3cm]{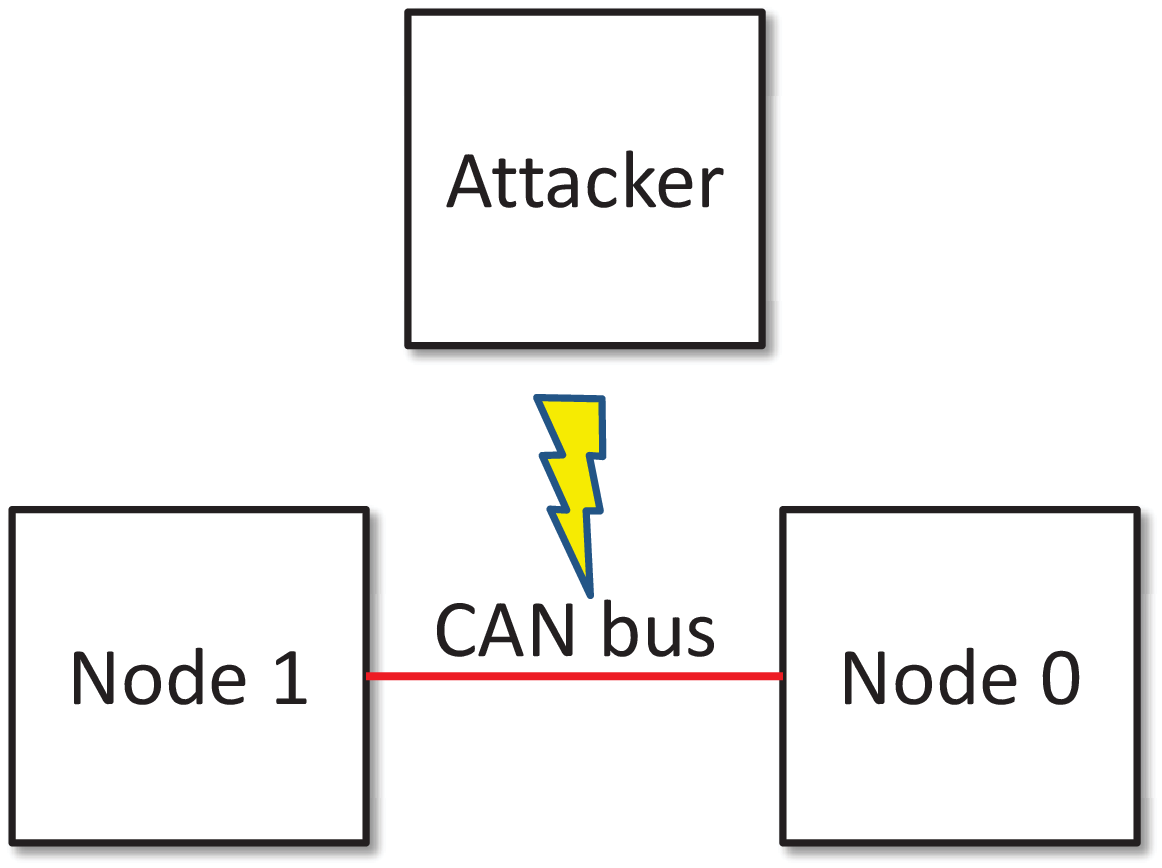}
\caption{DoS attacks on the CAN bus.}
\label{fig:two_adversary_nodes}
\end{subfigure}
\begin{subfigure}[t]{0.24\textwidth}
\centering
\includegraphics[scale=0.33, clip=true, trim=6.24cm 8.03cm 7.36cm 2.3cm]{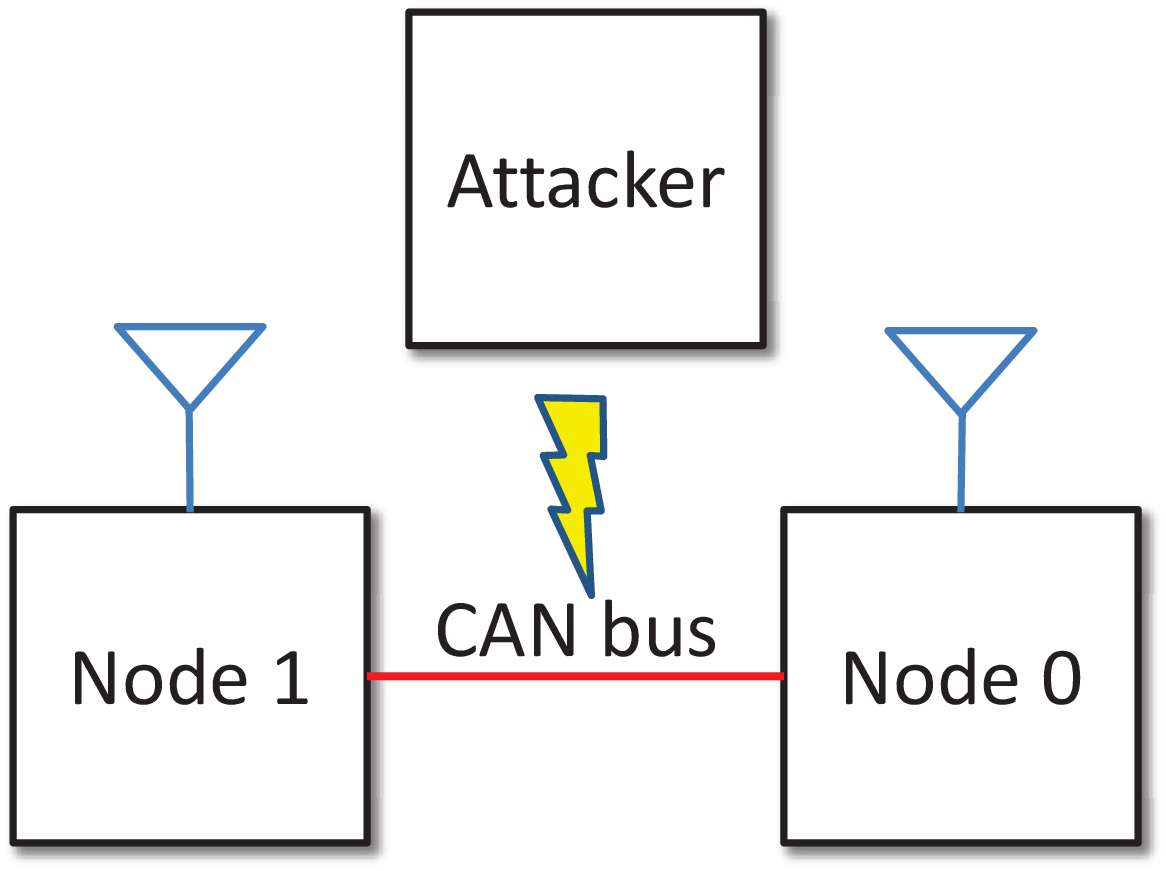}
\caption{DoS attacks on the hybrid network.}
\label{fig:two_adversary_nodes_hybrid}
\end{subfigure}
\begin{subfigure}[b]{0.24\textwidth}
\centering
\includegraphics[scale=0.33, clip=true, trim=6.24cm 5.25cm 7.36cm 2.5cm]{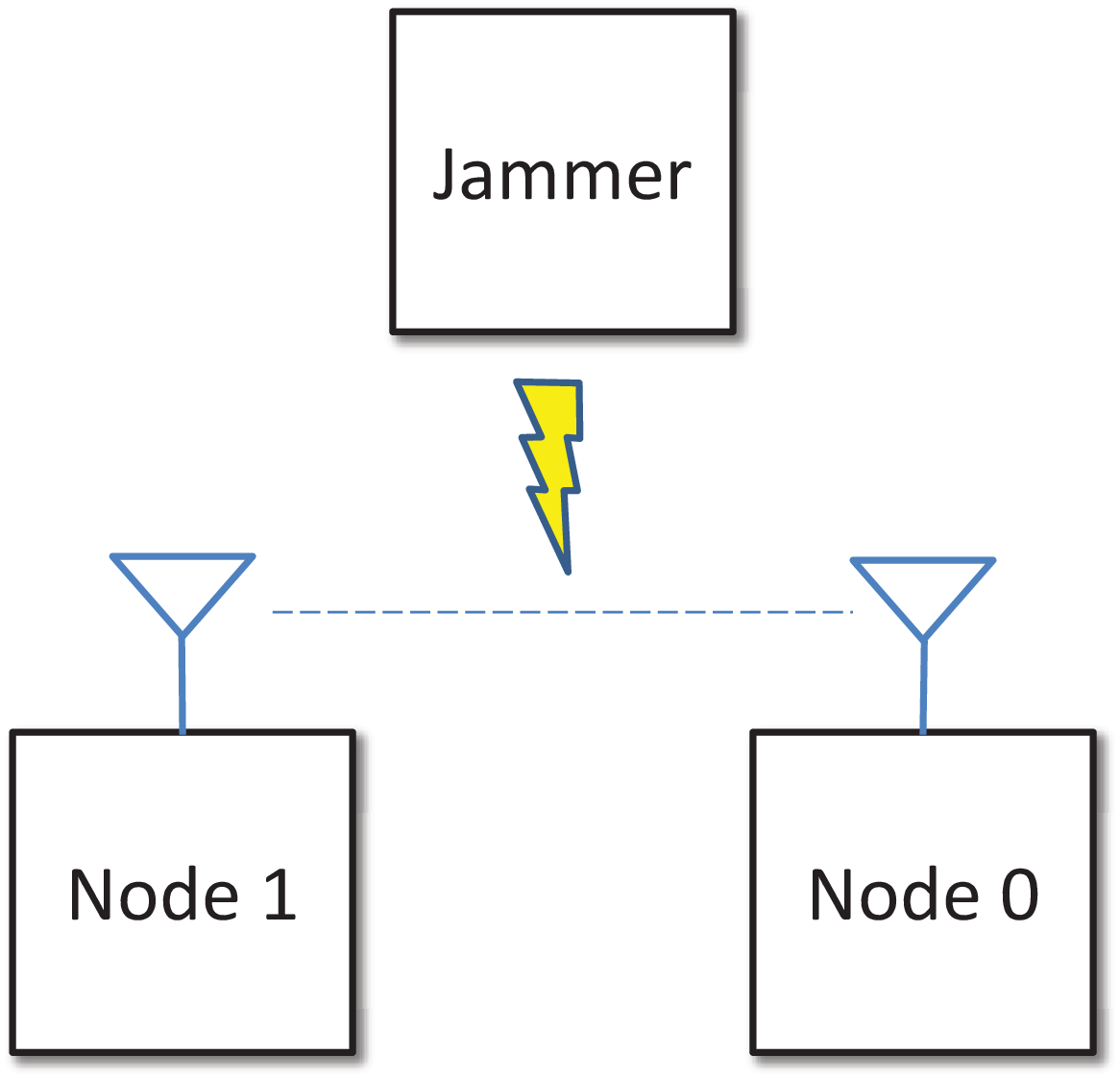}
\caption{Wireless jamming on the ZigBee link.}
\label{fig:one_wireless_jammer}
\end{subfigure}
\begin{subfigure}[b]{0.24\textwidth}
\centering
\includegraphics[scale=0.33, clip=true, trim=6.24cm 5.25cm 7.36cm 2.5cm]{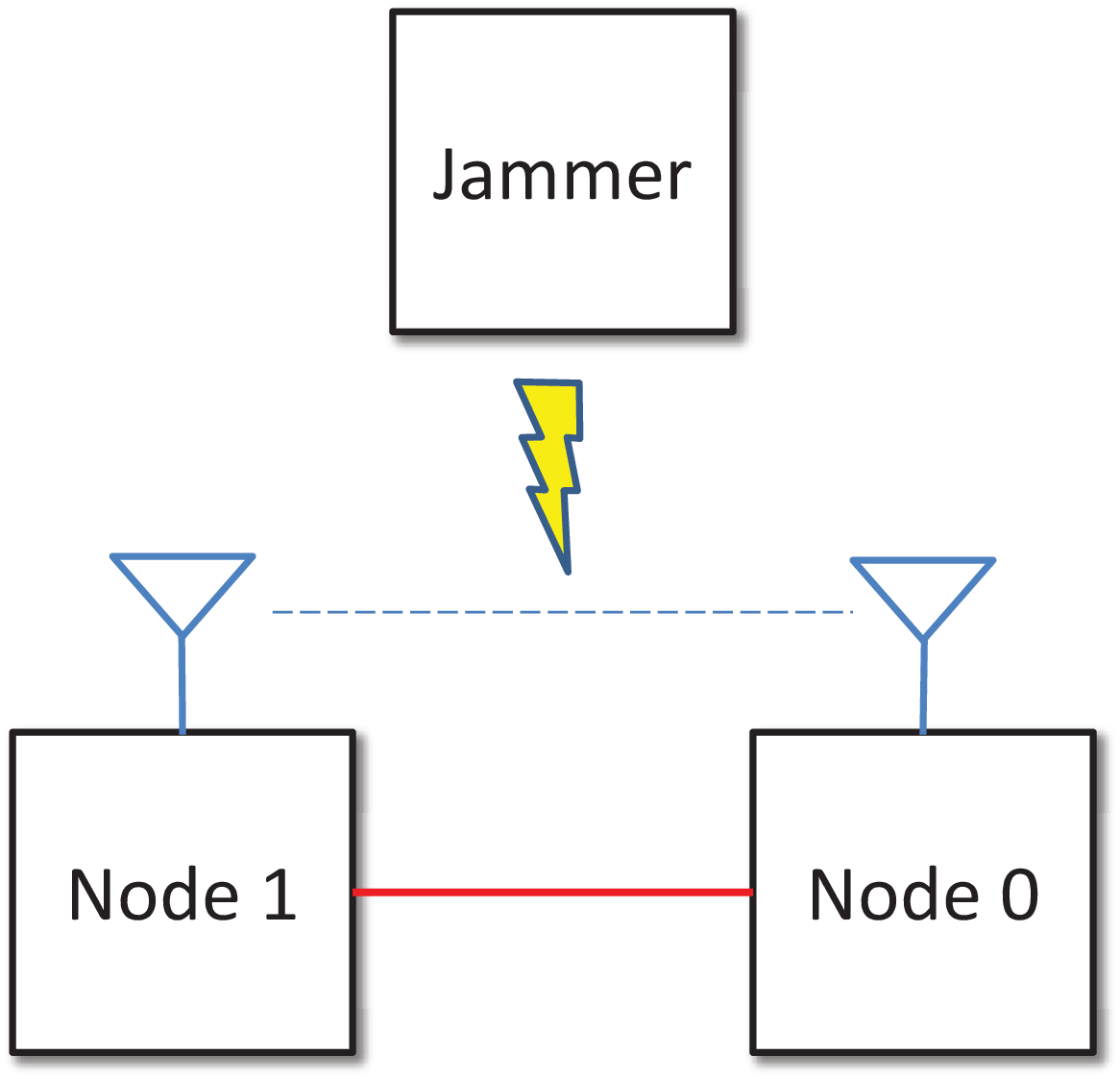}
\caption{Wireless jamming on the hybrid network.}
\label{fig:one_wireless_jammer_hybrid}
\end{subfigure}
\caption{DoS attacks on the CAN bus and wireless jamming attacks.}
\end{figure}
%
%

\begin{figure}[t!]
\begin{subfigure}[t]{0.48\textwidth}
\centering
\includegraphics[scale=0.4]{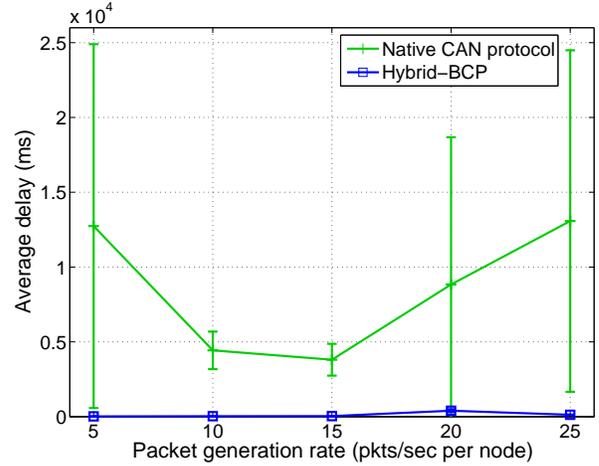}
\caption{Average delay of the legitimate node versus its packet generation rate.}
\label{fig:delay_two_jammer}
\end{subfigure}
\begin{subfigure}[t]{0.48\textwidth}
\centering
\includegraphics[scale=0.4]{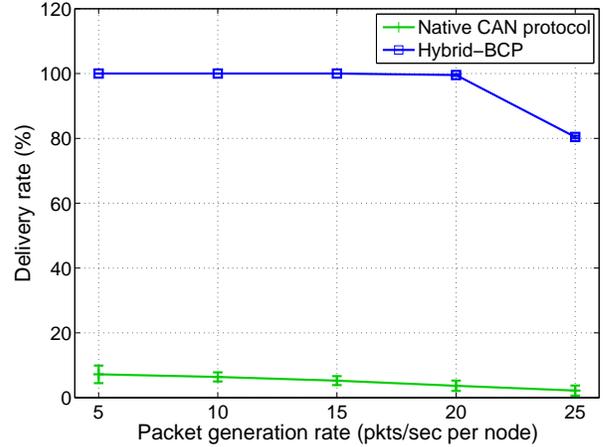}
\caption{Delivery rate of the legitimate node versus its packet generation rate.}
\label{fig:delivery_rate_two_jammer}
\end{subfigure}
\begin{subfigure}[t]{0.48\textwidth}
\centering
\includegraphics[scale=0.4]{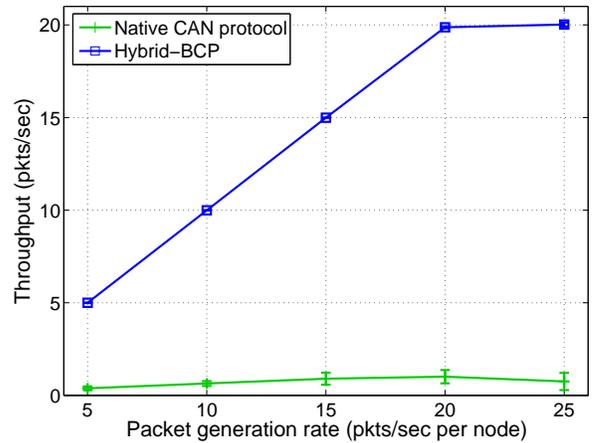}
\caption{Throughput of the legitimate node versus its packet generation rate.}
\label{fig:throughput_two_jammer}
\end{subfigure}
\caption{Performance of the native CAN protocol and Hybrid-BCP under DoS attacks on the CAN bus. 
}
\label{fig:performance_two_jammer}
\vspace*{-0.3cm}
\end{figure}

The CAN protocol is a priority-based protocol: a high-priority message gets transmitted ahead of a low-priority message. Previous work in \cite{Checkoway-2011} shows that the injection of malicious CAN messages can be done by remotely compromising and controlling nodes on the bus (via the radio, the tire pressure monitoring system or the Bluetooth component). 
In this section, we evaluate the effects of such DoS attacks on a \emph{legitimate node} that has not been compromised. 

We implement a DoS attack by having an attacker transmit high-priority messages on the CAN bus, at a rate of 300 pkts/sec. 
We compare Hybrid-BCP to the \emph{native CAN protocol}, a transmission scheme in which a legitimate node periodically generates data packets and transfers them to the CAN transceiver. The performance of the native CAN protocol is tested in the network of Fig. \ref{fig:two_adversary_nodes} and the performance of Hybrid-BCP is tested in the network of Fig. \ref{fig:two_adversary_nodes_hybrid}.

Fig. \ref{fig:delay_two_jammer} shows the average delay of packets by node 1 under the native CAN protocol and Hybrid-BCP. We see that the average delay of the native CAN protocol under the DoS attack reaches high values exceeding 3 sec (e.g., 3,810 ms at a packet generation rate of 15 pkts/sec by a legitimate node). 
The low-priority legitimate node is almost starved and must wait for a long time between each successful transmission. This result indicates that the DoS attack dramatically increases the MAC delay of a legitimate node.

\begin{figure}[t!]
\begin{subfigure}[t]{0.48\textwidth}
\centering
\includegraphics[scale=0.4]{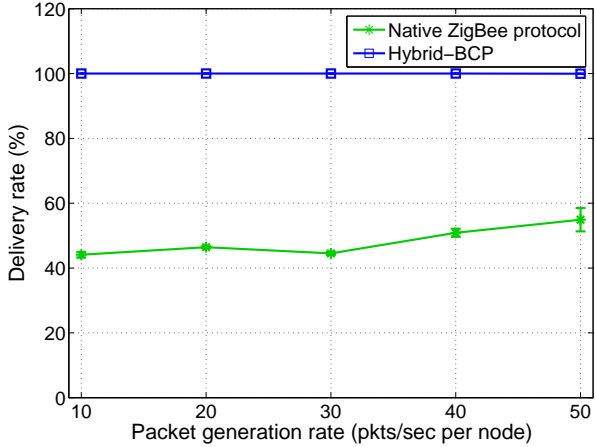}
\caption{Delivery rate of the legitimate node versus its packet generation rate.}
\label{fig:delivery_rate_one_wireless_jammer}
\end{subfigure}
\begin{subfigure}[t]{0.48\textwidth}
\centering
\includegraphics[scale=0.4]{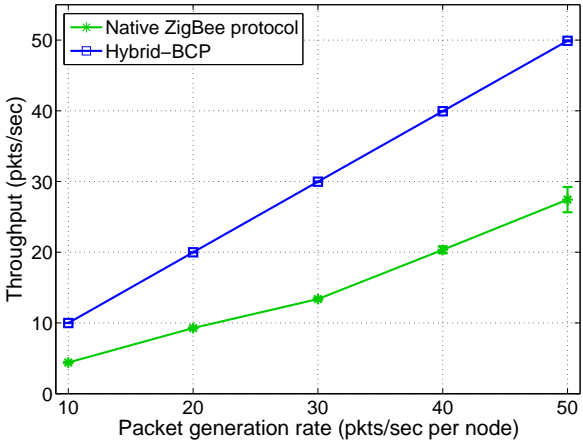}
\caption{Throughput of the legitimate node versus its packet generation rate.}
\label{fig:throughput_one_wireless_jammer}
\end{subfigure}
\caption{Performance of the native ZigBee protocol and Hybrid-BCP under wireless jamming attacks.}
\end{figure}

On the other hand, under the same DoS attack, Hybrid-BCP achieves higher packet delivery rate and higher throughput than the native CAN protocol, as shown in Fig. \ref{fig:delivery_rate_two_jammer} and \ref{fig:throughput_two_jammer}.
More specifically, Hybrid-BCP achieves a throughput of 19.87 pkts/sec at a packet generation rate of 20 pkts/sec by a legitimate node, more than ten times higher than that of the native CAN protocol. This gain is achieved because Hybrid-BCP measures a high ETX on the CAN link and schedules packet transmissions on the ZigBee link instead.
These results demonstrate that Hybrid-BCP can protect the CAN bus against DoS attacks.


\subsubsection{Wireless jamming}

In the wireless jamming tests, a wireless jammer performs protocol-compliant jamming attacks on the ZigBee link. The jammer periodically generates packets and broadcasts them on the ZigBee link. In our tests, the rate the wireless jammer generates packets is 100 pkts/sec. We compare Hybrid-BCP with the \emph{native ZigBee protocol}, which simply periodically generates data packets and sends them on the wireless link to the sink. The native ZigBee protocol is tested in the network of Fig. \ref{fig:one_wireless_jammer} and the hybrid wired/wireless protocol is tested in network of Fig. \ref{fig:one_wireless_jammer_hybrid}.

Comparisons between the delivery rate and throughput of the native ZigBee protocol and those of Hybrid-BCP are shown in Fig. \ref{fig:delivery_rate_one_wireless_jammer} and Fig. \ref{fig:throughput_one_wireless_jammer}. The results show that under wireless jamming, the delivery rate of the ZigBee protocol is at most 54.90\% at a packet generation rate of 50 pkts/sec, while Hybrid-BCP achieves a delivery rate of 99.95\%.

\section{Simulations}
\label{sec:simulations}

In this section, we provide performance evaluation of Hybrid-BCP in ns-3 simulations. We demonstrate the load balancing functionality of Hybrid-BCP under a higher wireless data rate. We also compare Hybrid-BCP to a tree-based routing protocol in a simulated intra-car hybrid wired/wireless network.


\subsection{Simulation setup}

\begin{table}[t!]
\centering
\begin{tabular}{|l|r|}
\hline
Ethernet data rate & 4Mbps \\ \hline
Wi-Fi standard & 802.11b \\ \hline
Wi-Fi mode & Ad hoc \\ \hline
Wi-Fi data rate & DSSS 11Mbps \\ \hline
Ethernet ACK timeout of Hybrid-BCP & 1 ms \\ \hline
Wi-Fi ACK timeout of Hybrid-BCP & 2 ms \\ \hline
\end{tabular}
\vspace{0.3cm}
\caption{The parameters in ns-3 simulations of Hybrid-BCP.}
\label{tab:ns3_simulation_parameters}
\end{table}

In the ns-3 simulations, we use the built-in Ethernet and Wi-Fi ns-3 libraries to simulate wired and wireless links. To emulate the CAN/ZigBee links in the real testbed, we configure the ns-3 simulations such that the Wi-Fi link has a higher rate but a larger RTT than the Ethernet link. Table \ref{tab:ns3_simulation_parameters} describes the simulation configuration and the parameters of Hybrid-BCP. The simulation configuration is used throughout the simulations except for Section \ref{sec:sim_load_balancing}, in which we compare the performance of Hybrid-BCP under different Wi-Fi rates. Each simulation is run for five times to obtain an average and a 95\% confidence interval for the metrics.





\begin{figure}[t!]
\captionsetup[subfigure]{labelformat=empty}
\centering
\begin{subfigure}[t]{0.24\textwidth}
\centering
\includegraphics[scale=0.35, clip=true, trim=6.45cm 9.69cm 7.26cm 5.61cm]{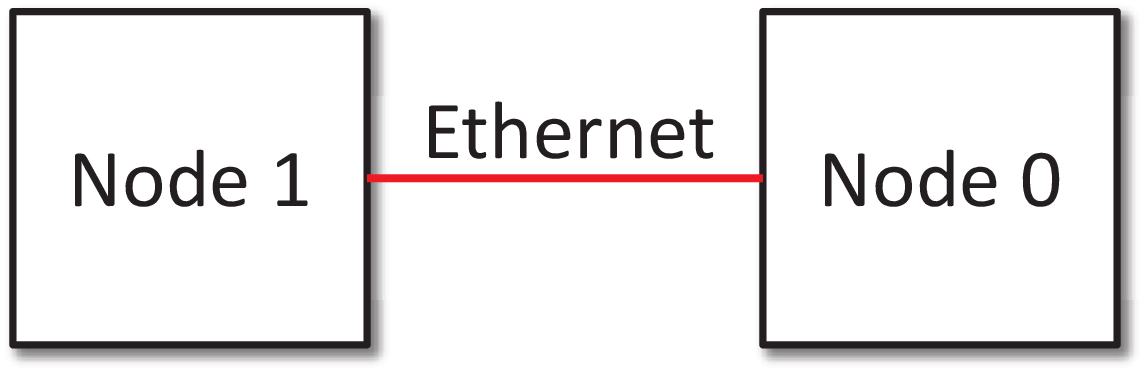}
\label{fig:two_node_wired_network}
\caption{Network D}
\end{subfigure}
\begin{subfigure}[t]{0.24\textwidth}
\centering
\includegraphics[scale=0.35, clip=true, trim=6.45cm 8.87cm 7.26cm 4.63cm]{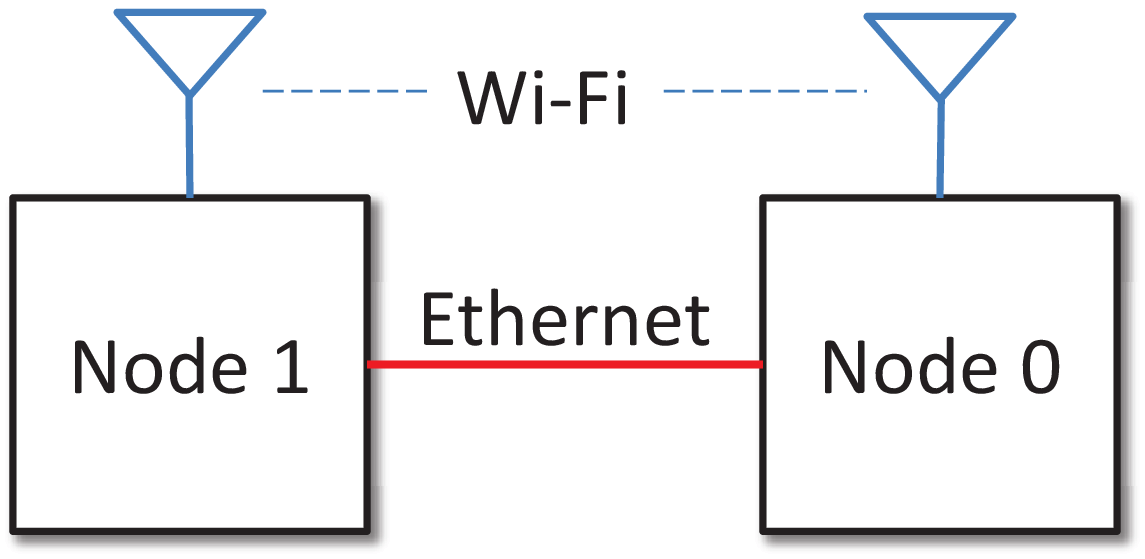}
\caption{Network E}
\end{subfigure}
\caption{The network topologies used for demonstrating the load balancing functionality of Hybrid-BCP in simulations.}
\label{fig:topologies_load_balancing_ns3}
\vspace*{-0.3cm}
\end{figure}

\subsection{Load balancing} \label{sec:sim_load_balancing}

\begin{figure}[t!]
\includegraphics[scale=0.4]{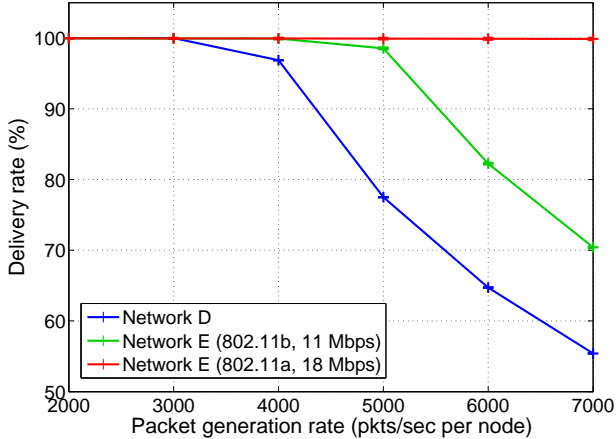}
\caption{Delivery rates of Hybrid-BCP on network D (Ethernet only) and network E (hybrid) in ns-3 simulations. The throughput improvement by load balancing of Hybrid-BCP is more significant as the wireless data rate gets higher.}
\label{fig:delivery_rate_3node_diff_wifi_rates}
\vspace*{-0.3cm}
\end{figure}

With the extra wireless link having a higher data rate, Hybrid-BCP can perform load balancing to aggregate more bandwidth. We run simulations of Hybrid-BCP in three scenarios: (1) network A; (2) network B with Wi-Fi rate equal to 11 Mbps (802.11b); (3) network B with Wi-Fi rate equal to 18 Mbps (802.11a). The packet delivery rates under the three scenarios are plotted in Fig. \ref{fig:delivery_rate_3node_diff_wifi_rates}. At a packet generation rate of 6,000 pkts/sec, Hybrid-BCP achieves a delivery rate of 61.71\% when there is no extra wireless link. With an extra Wi-Fi link at the rate of 11 Mbps, the delivery rate is increased to 82.25\%. The delivery rate is further increased to 99.90\% when the Wi-Fi data rate is increased to 18 Mbps. The results show that the benefits of load balancing by Hybrid-BCP are more significant as the wireless data rate gets higher.

\subsection{Routing and robustness}

\begin{figure}[t!]
\begin{subfigure}[t]{0.48\textwidth}
\centering
\includegraphics[scale=0.4]{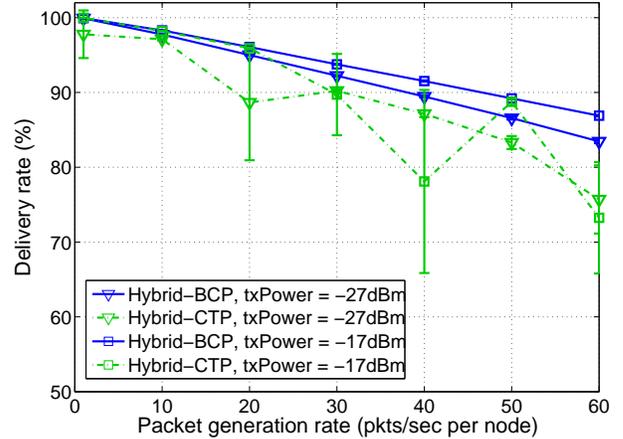}
\caption{Delivery rate versus packet generation rate.}
\label{fig:delivery_rate_intra-car_bcp_ctp}
\end{subfigure}
\begin{subfigure}[t]{0.48\textwidth}
\centering
\includegraphics[scale=0.4]{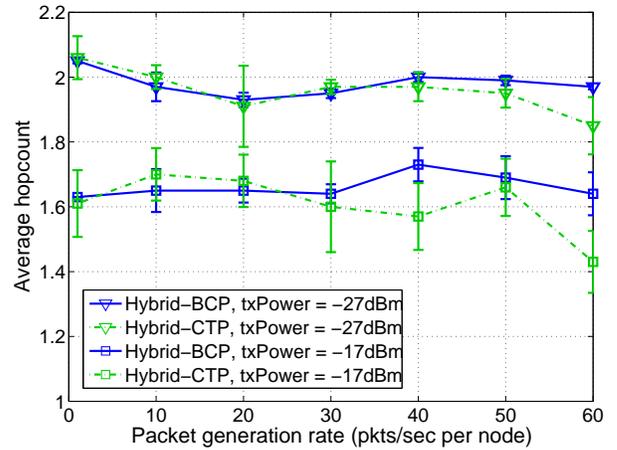}
\caption{Average hopcount versus packet generation rate.}
\label{fig:avg_hopcount_intra-car_bcp_ctp}
\end{subfigure}
\caption{Performance of Hybrid-BCP and Hybrid-CTP in a simulated 15-node intra-car hybrid wired/wireless network. Hybrid-BCP achieves comparable reliability with Hybrid-CTP even when reducing the radio power by 10 dBm.
}
\vspace*{-0.3cm}
\end{figure}


We use the intra-car RSSI traces measured from real experiments to simulate a 15-node intra-car hybrid wired/wireless network (the topology is shown in Fig. \ref{fig:hybrid_network_illustrated}). In the hybrid network, the sink is on the driver seat, three sensors are placed in the engine compartment, four sensors are respectively attached to the four wheels, three sensors are placed on passenger seats and the rest placed on the chassis. We use Hybrid-CTP, a variant of CTP designed for the hybrid network (see the appendix for description of the protocol), as a tree-based routing protocol to compare with Hybrid-BCP. In the simulations, each sensor periodically generates and transfers data packets to the underlying protocol (Hybrid-BCP or Hybrid-CTP), which routes the packets to the sink.


Fig. \ref{fig:delivery_rate_intra-car_bcp_ctp} and \ref{fig:avg_hopcount_intra-car_bcp_ctp} show the packet delivery rate and average hop count of Hybrid-BCP and Hybrid-CTP under two different radio power levels. Fig. \ref{fig:delivery_rate_intra-car_bcp_ctp} shows that when the radio transmission power is -27 dBm, Hybrid-BCP achieves higher packet delivery rate. At a packet generation rate of 20 pkts/sec, Hybrid-BCP achieves a delivery rate of 95\% while Hybrid-BCP achieves a delivery rate of 88.66\%. This shows that Hybrid-BCP outperforms Hybrid-CTP in the intra-car hybrid network and is thus more robust to the dynamic intra-car wireless links. 
When the radio transmission power is increased to -17 dBm, the delivery rate of Hybrid-CTP improves but is still lower than Hybrid-BCP at -27 dBm (e.g., at a packet generation rate of 30 pkts/sec). This indicates that Hybrid-BCP can consume less power to achieve the same reliability performance, and thus is more power efficient.


The routing functionality of Hybrid-BCP is further demonstrated by the statistics of number of hops in Fig. \ref{fig:avg_hopcount_intra-car_bcp_ctp}. The figure shows that when the radio transmission power increases, the average number of hops decreases, which is to our expectation.

\section{Conclusion} \label{sec:conclusion}

In this paper, we designed and implemented Hybrid-BCP as a joint load balancing and routing solution for data collection in intra-car hybrid wired/wireless networks. 
It is backward-compatible with existing intra-car wired network since no modification is needed on the CAN protocol.
We demonstrated the load balancing and routing functionalities of Hybrid-BCP in testbed experiments. We showed that Hybrid-BCP can be used to alleviate the impact of DoS attacks on the CAN bus.
Through simulations of intra-car hybrid networks based on dynamic RSSI traces collected from real experiments, we showed that Hybrid-BCP can use less radio transmission power to achieve the same reliability as the tree-based collection protocol.
Additional simulation results showed that the improvement on throughput by Hybrid-BCP is more significant with a wireless link of higher data rate.



For the sensors in the car, different sensors have different priority on transmitting their messages. 
It would be useful to extend Hybrid-BCP to incorporate a priority-based load balancing and routing scheme. The priority information needs to be incorporated in the protocol header and the protocol should be devised to reasonably make use of this information. It is also interesting to extend Hybrid-BCP to support data collection with multiple sinks, where each sensor is assigned a specific sink. We leave these as future works.

\appendix[Protocol design of Hybrid-CTP]

\begin{algorithm}[t!]
\caption{Hybrid-CTP}
\begin{varwidth}{\dimexpr\linewidth-2\fboxsep-2\fboxrule\relax}
\begin{algorithmic}[1]
\Procedure{Wired\_interface\_handler}{}
\While{$\mathcal{Q}_i > 0$}
\State Compute the $ETX_{i,j}^{W}$ for each neighbor $j$ on \hspace*{1.05cm}the wired link
\State Find the neighbor $j^*$ such that $j^*=$ \hspace*{1.05cm}$\operatorname*{arg\,min}_{j} ETX_{i,j}^{W}$
\If {$ETX^{W^*}_{i} < ETX^{WL^*}_i + T$}
\State Transmit one packet to $j^*$ over the wired \hspace*{1.57cm}interface
\State Update $\overline{ETX}^{W}_{i\rightarrow j^*}$ and $ETX_i$
\Else
\State Wait for a reroute period 
\EndIf
\EndWhile
\EndProcedure
\State 	
\Procedure{Wireless\_interface\_handler}{}
\While{$\mathcal{Q}_i > 0$}
\State Compute the $ETX_{i,k}^{WL}$ for each neighbor $k$ \hspace*{1.05cm}on the wireless link
\State Find the neighbor $k^*$ such that $k^*=$ \hspace*{1.05cm}$\operatorname*{arg\,min}_{k} ETX_{i,k}^{WL}$
\If {$ETX^{WL^*}_{i} < ETX^{W^*}_i + T$}
\State Transmit one packet to $k^*$ over the wireless \hspace*{1.57cm}interface
\State Update $\overline{ETX}^{WL}_{i\rightarrow k^*}$ and $ETX_i$
\Else
\State Wait for a reroute period 
\EndIf
\EndWhile
\EndProcedure
\end{algorithmic}
\end{varwidth}%
\end{algorithm}

In this section, we describe Hybrid-CTP, a variant of CTP designed for data collection in hybrid wired/wireless networks.

The same as Hybrid-BCP, Hybrid-CTP has two procedures handling the wired and wireless interfaces, respectively. Suppose for node $i$, node $j$ is a neighbor on interface $I$, where $I \in \{W, WL\}$ ($W$ represents the wired interface and $WL$ represents the wireless interface). 
Let $\overline{ETX}^I_{i\rightarrow j}$ denote an estimate of the average number of transmissions needed to successfully transmit a packet from $i$ to $j$ over interface $I$.


Each node $i$ records its end-to-end \emph{path cost} to the sink, denoted by $ETX_i$. To obtain $ETX_i$, 
node $i$ first calculates the path cost through interface $I$ via node $j$ as follows:
\[
ETX^{I}_{i,j}= ETX_{j} + \overline{ETX}^{I}_{i\rightarrow j}.
\]
The minimum path cost from node $i$ to the sink node through interface $I$ is $ETX^{I^*}_{i} = \min_j ETX^{I}_{i,j}$.


Then node $i$ calculates its path cost to the sink by:
\[
ETX_i = \min \{ETX^{W^*}_i, ETX^{WL^*}_i\}.
\]

The path cost information is propagated to neighbors by beacon messages, the same as the backpressure information in Hybrid-BCP. The sink broadcasts path cost of zero. 

In Hybrid-CTP, the wired interface handler schedules a packet transmission when $ETX^{W^*}_{i} < ETX^{WL^*}_i + T$, where $T$ is a positive integer (set to two in our implementation). Similarly, the wireless interface handler schedules a packet transmission when $ETX^{WL^*}_{i} < ETX^{W^*}_i + T$. Therefore, when $ETX^{W^*}_i$ is much smaller than $ETX^{WL^*}_i$, only the wired interface handler will schedule packet transmissions. This could happen either when the wireless link quality is bad or when all the neighbors on the wireless link select this node as their next hop. When $ETX^{W^*}_i$ and $ETX^{WL^*}_i$ are close to each other, both interface handlers will transmit packets. Algorithm 3 provides a pseudo-code of Hybrid-CTP.

\bibliographystyle{ieeetr}
\bibliography{VNC_2015}  

\end{document}